\pdfoutput=1
\documentclass[iop,apj]{emulateapj}
\usepackage{graphicx,epsfig}
\usepackage{rotate}
\usepackage{amsmath, amsthm, amssymb}
\usepackage{times}
\usepackage{soul}
\usepackage{natbib}
\usepackage[plainpages=false, colorlinks=true, anchorcolor=blue,
linkcolor=blue, citecolor=blue, bookmarks=false]
{hyperref}

\newcommand{\bq}{\begin{equation}}
\newcommand{\eq}{\end{equation}}

\def\gtsim{\lower.5ex\hbox{$\buildrel > \over\sim$}}
\def\ltsim{\lower.5ex\hbox{$\buildrel < \over\sim$}}

\def\sun{\mbox{$_\odot$}}
\def\apjl{ApJL}

\def\apjs{ApJS}

\def\aap{A\&A}

\newcommand{\subdate}{2016 January 21}
\newcommand{\shortauth}{Chatzopoulos et al.}
\newcommand{\slugcom}{Submitted to ApJ on \subdate}
\slugcomment{\slugcom}

\lefthead{\sc \footnotesize \slugcom \hfill \shortauth}
\righthead{\sc \footnotesize \slugcom \hfill \shortauth}
\begin{document}
\title
{Convective Properties of Rotating Two-dimensional Core-collapse Supernova Progenitors}
\author{E. Chatzopoulos\textsuperscript{1,}\altaffilmark{9}}
\author{Sean M. Couch\textsuperscript{2,3,4,5}}
\author{W. David  Arnett \textsuperscript{6,7}}
\author{F. X. Timmes \textsuperscript{8}}
\affil{
  \altaffilmark{1}{Department of Astronomy \& Astrophysics, Flash Center for Computational
Science, University of Chicago, Chicago, IL, 60637, USA} \\
  \altaffilmark{2}{Department of Physics and Astronomy, Michigan State University, East Lansing, MI 48824, USA}\\
  \altaffilmark{3}{Department of Computational Mathematics, Science, and Engineering, Michigan State University, East Lansing, MI 48824, USA}\\
  \altaffilmark{4}{National Superconducting Cyclotron Laboratory, Michigan State University, East Lansing, MI 48824, USA}\\
  \altaffilmark{5}{Joint Institute for Nuclear Astrophysics, Michigan State University, East Lansing, MI 48824, USA}\\
  \altaffilmark{6}{Steward Observatory, University of Arizona, Tucson, AZ 85721, USA}\\
  \altaffilmark{7}{Aspen Center for Physics, Aspen, CO 81611, USA}\\
  \altaffilmark{8}{School of Earth and Space Exploration, Arizona State University, Tempe, AZ 85287, USA}
}
\altaffiltext{9}{Enrico Fermi Fellow; \href{mailto:manolis@flash.uchicago.edu}{manolis@flash.uchicago.edu}}

\begin{abstract}

We explore the effects of rotation on convective carbon, oxygen, and silicon 
shell burning during the late stages of evolution in a 20~$M_\odot$ star.
Using the Modules for Experiments in Stellar Astrophysics (MESA) we construct 1D stellar models both with no rotation and with an initial rigid rotation of 50\% of critical. 
At different points during the evolution, we map the 1D models into 2D and follow the multidimensional evolution using the FLASH compressible hydrodynamics code for many convective turnover times until a quasi-steady state is reached.
We characterize the strength and scale of convective motions via decomposition of the momentum density into vector spherical harmonics.
We find that rotation influences the 
total power in solenoidal modes, with a slightly larger impact for carbon and oxygen shell
burning than for silicon shell burning. 
Including rotation in one-dimensional (1D) stellar evolution models alters the structure of the star in a manner that has a significant impact on the character of multidimensional convection.
Adding modest amounts of rotation to a stellar model that ignores rotation during the evolutionary stage, however, has little impact on the character of resulting convection.
Since the spatial scale and strength of convection present at the point of core collapse directly influence the supernova mechanism, our results suggest that rotation could play an important role in setting the stage for massive stellar explosions.

\end{abstract}

\vskip 0.57 in

\section{INTRODUCTION}\label{intro}

The final years in the lives of massive stars are characterized by vigorous convective shell burning,
hydrodynamic and convective instabilities and, in many cases, episodic mass-loss events 
\citep{2007ApJ...667..448M,2012MNRAS.423L..92Q,2014ApJ...780...96S,2014ApJ...785...82S,2014AIPA....4d1010A} 
that change their three-dimensional 
structure and the initial conditions (ICs) for the core--collapse supernova (CCSN) explosion. 
In addition, rotation and magnetic fields may further complicate the 
core-collapse process in a non-linear fashion. 

The advanced stages of burning in massive stars have long been studied with 1D approximations 
\citep{1984psen.book.....C,2002RvMP...74.1015W}. 
For convection, the mixing-length theory (MLT;
\citealt{1958ZA.....46..108B}) remains the
technique used and implemented in stellar evolution codes with options to choose the 
associated efficiency parameter, $\alpha_{\rm MLT}$, and the condition determining where convection 
becomes active by using the Schwarzchild or the Ledoux criterion accounting for suppression 
due to compositional gradients. Parametrizations for 1D models of convective overshoot, semiconvection,
and thermohaline mixing are frequently adopted, as is the 1D shellular approximation to treat
rotation \citep{1992A&A...265..115Z, 1997A&A...321..465M}.

Advances in numerical algorithms, hydrodynamic software instruments, and computing power
have allowed multi--dimensional studies of stellar convection that can assess the fidelity of 1D
treatments of mixing.
The properties of convective oxygen shell--burning in the progenitor of SN~1987A prior to collapse
were studied by \citet{1998ApJ...496..316B} in 2D hydrodynamics simulations. One of the implications
of this study was the potential for
post-explosion mixing of radioactive $^{56}$Ni throughout the SN envelope. 
The first three-dimensional (3D) simulations of massive star oxygen shell convection were 
presented by \citet{2006PhDT........20M} and \citet{2007ApJ...667..448M}.
This work showed that the boundaries between non-convective and convective
regions are not stationary as standard MLT theory predicted but 
dynamical and 
the source
of gravity waves. \citet{2011ApJ...733...78A}  followed up with 2D simulations of simultaneously active
C, Ne, O and Si burning shells for a 23~$M_{\odot}$ progenitor 1~hr prior to core--collapse finding significant
departures from spherical symmetry and strong dynamical interactions between shells. 
The 3D hydrodynamics simulations together with developments in mathematical methods,
indicate that MLT needs significant revision especially for late stages of stellar evolution \citep{2015ApJ...809...30A,2015arXiv150505756G}.

The apparent departures from spherical symmetry that arise in the velocity fields of convective shells
prior to iron core--collapse have a qualitative and quantitative impact to the susceptibility to explosion itself \citep{2013ApJ...778L...7C, 2015ApJ...799....5C, 2015MNRAS.448.2141M}.
\citet{2013ApJ...778L...7C, 2015ApJ...799....5C} have shown that successful explosions occur
for models whose ICs include velocity field perturbations due to
convective burning, 
in contrast to the same models without these asphericities. 
This is due to
non-radial motions 
in the accretion flow exciting stronger post-shock turbulence that aids shock revival.
Full 3D simulations of the final minutes of
iron core growth and collapse also 
suggest that non-spherical progenitor structure
should not be ignored \citep{2015ApJ...808L..21C}. 
One might
characterize convection seen in 2D or 3D hydrodynamic simulations and then map realizations of 
convective velocity fields as ICs for simulations of CCSNe. Efforts to 
do so
include the spherical Fourier--Bessel decomposition \citep{2014MNRAS.440.2763F} 
and nonlocal and time-dependent averaging \citep{2007ApJ...667..448M,2009ApJ...690.1715A,2013ApJ...769....1V}.
In our analysis we use vector spherical harmonic (VSH) decomposition of the momentum density
field as presented by \citet{2014ApJ...795...92C}.

The stochastic nature of 1D convection algorithms
 has a profound impact to the outcome of massive stellar evolution itself.
For stars of very similar mass and metallicity the end points can be dramatically different. 
These give remnant masses and explosion properties which depend strongly on the pre--SN stellar structure and exhibit large
variability even in narrow intervals of Zero Age Main Sequence (ZAMS) mass \citep{2012ApJ...757...69U, 2014ApJ...783...10S}. 

Additional uncertainty arises from 
the dynamical boundaries between convective zones \citep{2007ApJ...667..448M},
which 
are a source of gravity waves
and might lead to episodic mass-loss events shortly before the SN explosion
\citep{2012MNRAS.423L..92Q,2014ApJ...780...96S,2014ApJ...785...82S}. The circumstellar (CS) environment that is formed
around pre--SN stars due to this mass--loss history can have a significant impact on the radiative properties
of the resulting explosion. 

\begin{figure}
\begin{center}
\includegraphics[angle=0,width=6cm,trim=0.1in 0.in 0.1in 0.in,clip]{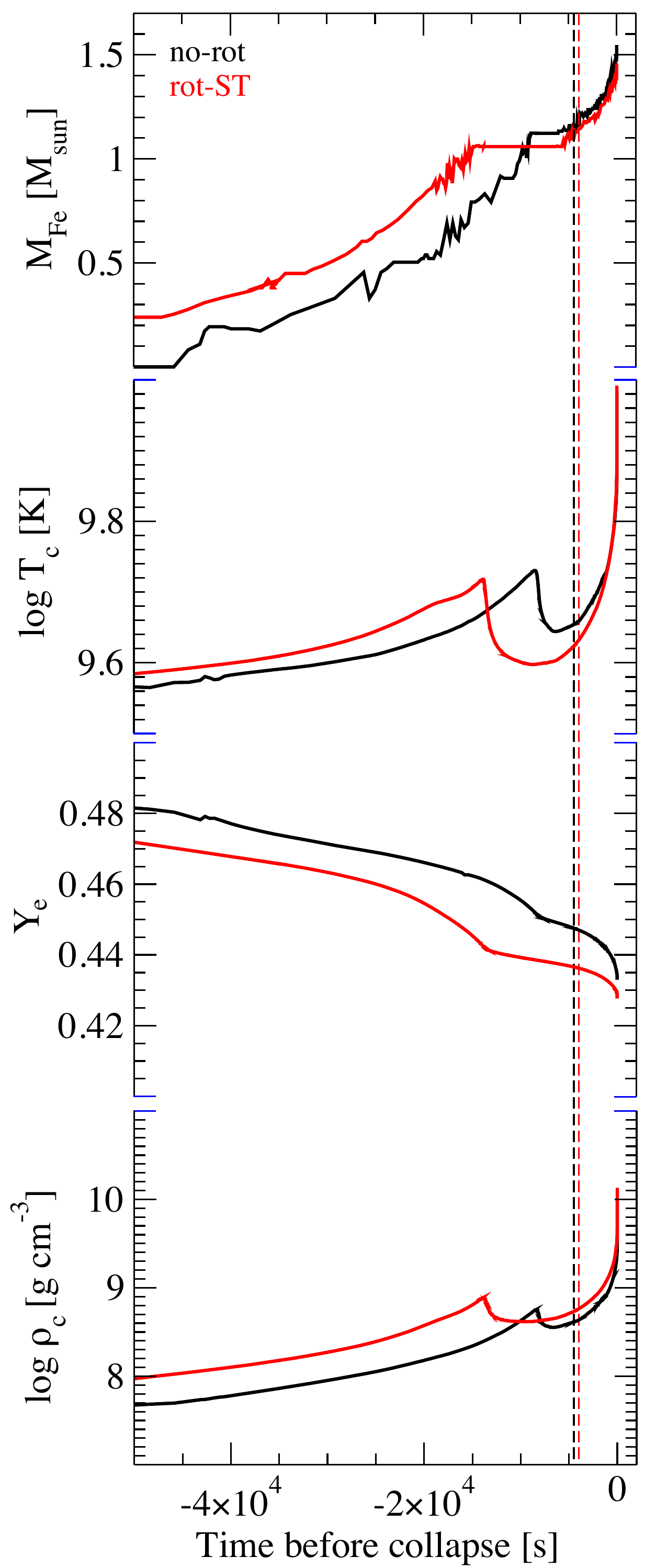}
\caption{From top to bottom: evolution of global parameters for the ``no-rot'' (solid black curves) and the ``rot-ST''
(solid red curves) models as calculated in {\it MESA}. Iron core mass ($M_{\rm Fe}$),
central temperature ($T_{\rm c}$), central $Y_{\rm e}$ and
central density ($\rho_{\rm c}$). The dashed vertical lines indicate the times
that correspond to the Si-shell burning models mapped in {\it FLASH}.\label{Fig:evolution}}
\end{center}
\end{figure}

Rapid rotation has been observed for many massive stars 
\citep{2008ApJ...676L..29H, 2010A&A...512L...7V, 2011ApJ...743L..22D, 2015arXiv150908940A} and can affect their pre--SN internal structure
and composition via instabilities that alter the efficiency of angular momentum transport and the rate of chemical mixing
(\citealt{2011A&A...530A.115B, 2011A&A...530A.116B, 2012A&A...537A.146E}; see \citealt{2012RvMP...84...25M} for a review).
The effects of rotation on convection in the extended envelopes of red giant stars has been studied with 3D simulations by
\citet{2009ApJ...702.1078B}, who found the properties of turbulent convection to be sensitive to the rotation rate.
Currently, the effects of rotation on the convective properties of massive, pre--SN stars and their implications for CCSNe
have not been thoroughly investigated. In this paper we present 2D simulations of rotating pre--SN stars during convective
C-- and O-- shell burning and Si-- shell burning, and use the VSH method to quantify the behavior of convective velocity flows 
as a function of internal rotation rate. 

The paper is organized as follows. In \S~\ref{Stellevol} we present the stellar evolution and 2D hydrodynamics simulations
of a 20~$M_{\odot}$ pre--SN star during convective shell burning of C, O and Si. In \S~\ref{VSHdecomp} we apply the
method of VSH to decompose the convective velocity fields and obtain power spectra in order to characterize the
convective properties of models of different rotation rates. Finally, in \S~\ref{Disc} we discuss our conclusions and
implications for CCSNe. 

\section{STELLAR EVOLUTION AND HYDRODYNAMICS SIMULATIONS}\label{Stellevol}

Our analysis of 2D convection in rotation pre--SN progenitors has three distinct steps.
First, we evolve models of a star with ZAMS mass of 20~$M_{\odot}$, solar metallicity but different rotation rates using
version 7503 of the stellar evolution code Modules for Experiments in Stellar Astrophysics 
({\it MESA}; \citealt{2011ApJS..192....3P, 2013ApJS..208....4P, 2015ApJS..220...15P}). Subsequently, 
{\it MESA} 1D profiles are extracted days prior to core--collapse, during core Si-- and shell C-- and O--burning, and
$\sim$~1~hr prior to collapse during Si--shell burning.
These profiles are then mapped into 2D using the adaptive mesh refinement (AMR) multi-physics hydrodynamics code {\it FLASH} version 4.3
\citep{2000ApJS..131..273F, 2012ApJS..201...27D} including rotation perpendicular to the plane of
the simulation (``2.5D'' approach). The 2D simulations are run for $>$~3 convective turn-over timescales, long enough to
diminish the effects of the initial dynamical transient resulting from mapping 1D 
``convective" profiles to a multi-dimensional
hydrodynamic grid. Lastly, {\it FLASH} simulation output at three different times is extracted and post-processed 
using the VSH method \citep{2014ApJ...795...92C} to obtain the power spectra of the convective motions and the energy cascade. 

\subsection{{\it MESA pre--SN evolution.}}\label{MESApreSN}

Our {\it MESA} models use initial rigid rotation rates of zero (``no--rot'') and 50\% of the critical Keplerial value (``rot--ST''). The initial rotation
profiles impose rigid--body rotation when the model first lands on the ZAMS. The rotating model assumes the transport of angular
momentum and chemical mixing via the Spruit--Taylor mechanism (ST; \citealt{1999A&A...349..189S, 2002A&A...381..923S}).

For MLT convection in {\it MESA} we adopt the Schwarzchild criterion and $\alpha_{\rm MLT}=$~1.6. We use an automatically
extending nuclear reaction network starting from a basic 8--isotope network and reaching a 21--isotope network 
by the end of the run. We note $\approx$~100-150 isotopes 
are required to accurately represent core neutronization and neutrino
cooling \citep{2011ApJ...733...78A, 2015ApJ...809...30A}. 
The ``Helmholtz'' equation of state (EOS;  \citealt{2000ApJS..126..501T}) 
is used. Standard mass--loss prescriptions appropriate for massive stars 
are adopted \citep{2001A&A...369..574V, 2009A&A...497..255G}. 
After performing a resolution study, we choose a spatial resolution parameter 
({\tt mesh\_delta\_coeff} in {\it MESA} terminology) equal to 0.5 and a temporal resolution factor
({\tt varcontrol\_target}) of $10^{-3}$ where good convergence ($\simeq$~10$^{-2}$ level) in terms
of final carbon-oxygen core mass and iron core mass is achieved. The chosen
grid resolution resulted in final output models with 1200--1600 Lagrangian zones. 

\begin{figure}
\begin{center}
\includegraphics[angle=0,width=7cm,trim=0.1in 0.in 0.1in 0.in,clip]{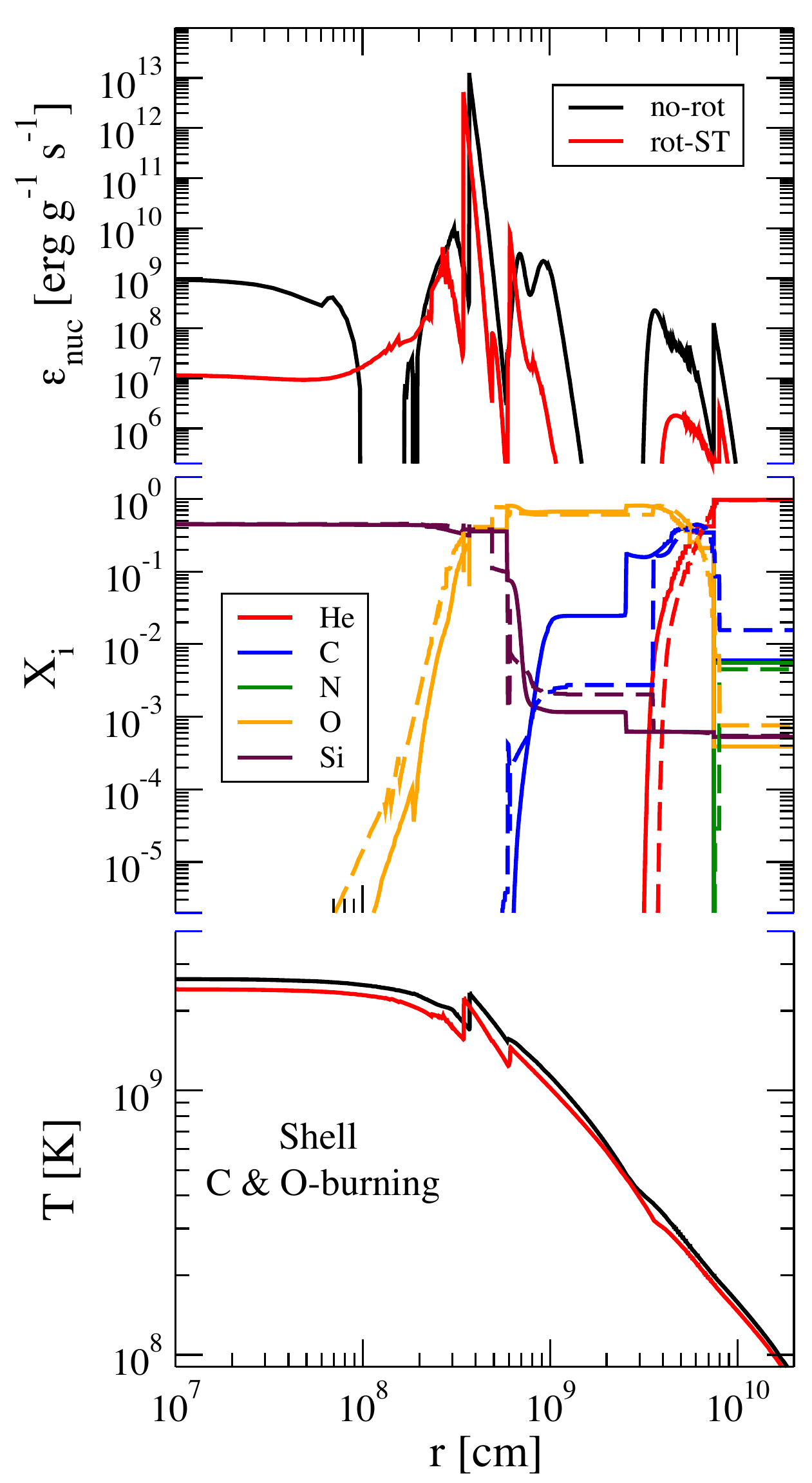}
\caption{From top to bottom: radial profiles of nuclear energy generation rate ($\epsilon_{\rm nuc}$), composition, 
and temperature ($T$) for the ``no-rot'' (solid black curves) 
and ``rot-ST'' (solid red curves) {\it MESA} models at the C\&O shell burning phase prior to mapping to 
{\it FLASH}. In the composition plot the ``no-rot'' model is represented by solid curves and the ``rot-ST'' model
by dashed curves. The vertical dashed lines denote the radial limits, $R_{\rm in,sh}$ and $R_{\rm out,sh}$ of the convective
shell that was decomposed with the VSH method.\label{Fig:CO_characteristics}}
\end{center}
\end{figure}

\begin{figure}
\begin{center}
\includegraphics[angle=0,width=7cm,trim=0.1in 0.in 0.1in 0.in,clip]{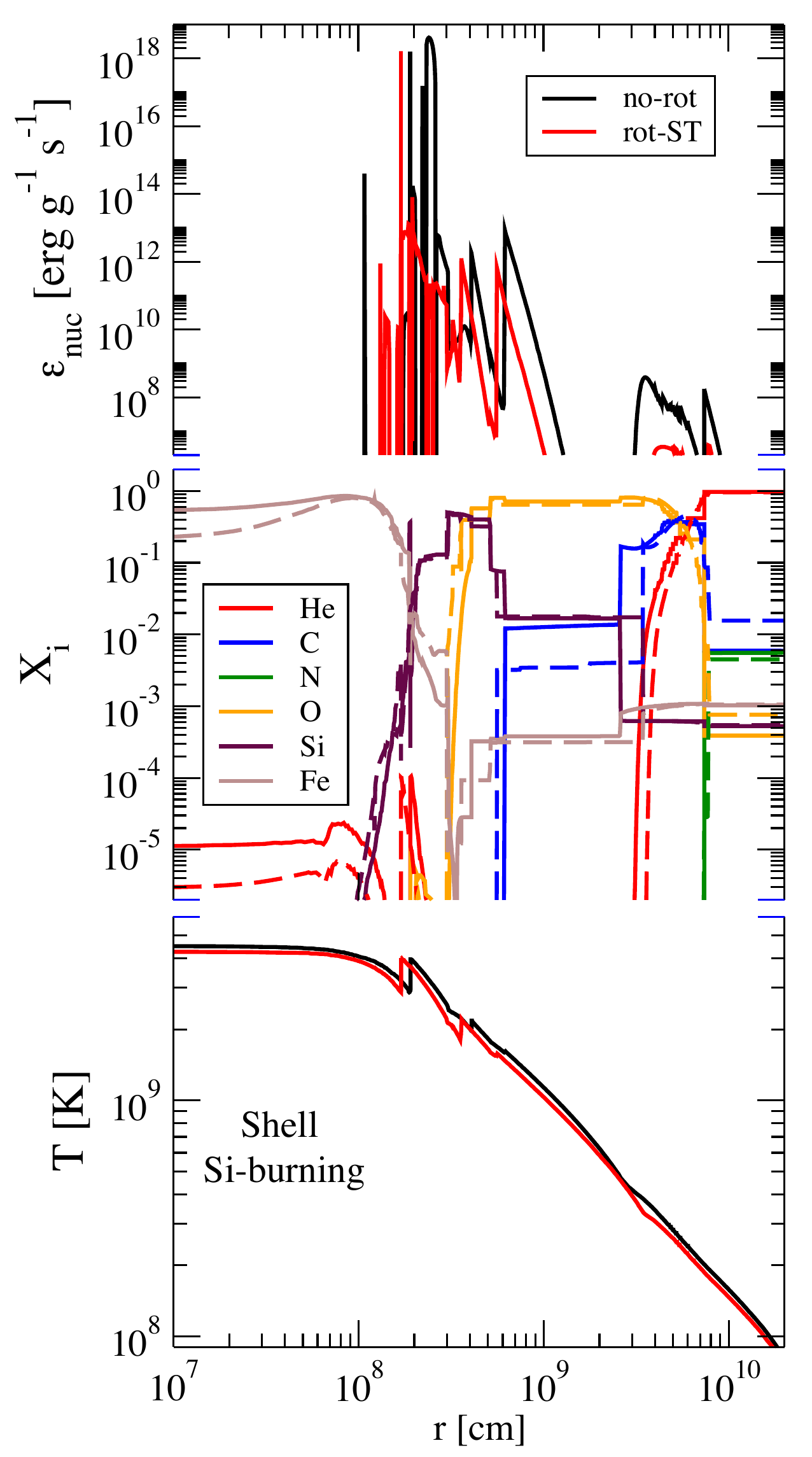}
\caption{Same as Figure~\ref{Fig:CO_characteristics} but for the Si shell burning phase.\label{Fig:Si_characteristics}}
\end{center}
\end{figure}

Figure~\ref{Fig:evolution} shows the evolution of iron core mass ($M_{\rm Fe}$) and central density ($\rho_{\rm c}$),
temperature ($T_{\rm c}$) and average charge per baryon ($Y_{\rm e}$) for both the ``no--rot'' and ``rot--ST'' models
for the last 50,000~sec of evolution and the formation of the iron core. The vertical dashed lines show the stage
where we extracted the {\it MESA} models corresponding to Si--shell burning, at 
$\sim$~1.3~hours prior to core--collapse, when the peak of $\epsilon_{\rm nuc}$
profile reached the maximum value ($3-4 \times 10^{18}$~erg~g$^{-1}$~s$^{-1}$).
The build--up of the Fe--core up to
the Chandrasekhar mass is smoother for the ``rot--ST'' model predominantly due to the effects of ST and rotational mixing. 

Figures~\ref{Fig:CO_characteristics} and ~\ref{Fig:Si_characteristics} show the distributions of nuclear energy
generation rate ($\epsilon_{\rm nuc}$), composition ($X_{\rm i}$)  and
temperature focused in the convective, shell-burning regions. 
The {\it MESA} models we calculate do not exhibit a phase of strongly
evident shell Ne--burning, in contrast to the the more massive (23~$M_{\odot}$) model used in \citet{2011ApJ...733...78A}.
The vertical dashed lines indicate the inner and outer radial
boundaries ($R_{\rm in,sh}$ and $R_{\rm out,sh}$) that were chosen for the convective shells to be analyzed with
the VSH method (see \S~\ref{VSHdecomp} for details on how their values were determined). 
The effects of enhanced mixing in the ``rot--ST'' model are clearly illustrated in the composition panels. 

\begin{figure}
\begin{center}
\includegraphics[angle=0,width=9cm,trim=0.4in 0.25in 0.5in 0.15in,clip]{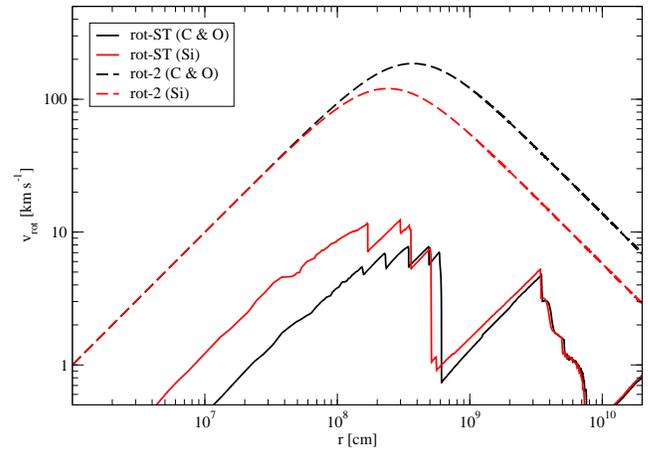}
\caption{Rotational velocity profiles for the ``rot-ST'' (solid curves) and the ``rot-2'' (dashed
curves) models. Black curves denote the C\&O shell burning phase and red curves the Si shell
burning phase. The form of the rotational profile for the ``rot-2'' models is given by Equation~\ref{Eq:rot_profile}.
\label{Fig:vrotprof}}
\end{center}
\end{figure}

To further isolate the effects of rotation, we impose an artifical rotational velocity
profile on the ``no--rot'' models:
\begin{equation}
v_{\rm rot}(r) = \frac{r \Omega}{1 + (r/A)^{2}},\label{Eq:rot_profile}
\end{equation}
where $r$ is the radial (spherical) coordinate, $\Omega$ is the angular velocity and $A$ the characteristic radius where
the rotational velocity peaks as in \citet{2013ApJ...776..129C}. 
We denote models that use this rotation law as ``rot--2''.
We carefully selected the values of $A$ and $\Omega$ in order to
capture a few full rotations (4--8) for the convective shell material 
within the simulated timescales in {\it FLASH} and assess the dynamical effects of the centrifugal force.
For C \& O--shell burning we picked $A = 3.46 \times 10^{8}$~cm and $\Omega =$~0.1~s$^{-1}$ and for
Si--shell burning $A = 2.2 \times 10^{8}$~cm and $\Omega =$~0.1~s$^{-1}$ respectively. 
Figure~\ref{Fig:vrotprof} shows the rotational velocity profiles for all models mapped in {\it FLASH}. 

\setcounter{table}{0}
\begin{deluxetable*}{lcccccccccc}
\tablewidth{0pt}
\tablecaption{Properties of MESA models mapped in {\it FLASH}.}
\tablehead{
\colhead {Model} &
\colhead {$R_{\rm in,sh}$~($10^{8}$~cm)} &
\colhead{$R_{\rm out,sh}$~($10^{8}$~cm)} &
\colhead {$v_{\rm rot,sh}$~(km~s$^{-1}$)} &
\colhead{$v_{\rm conv}$~(km~s$^{-1}$)} &
\colhead{$\tau_{\rm rot}$~(s)} &
\colhead {$t_{\rm 1}$~(s)} &
\colhead {$t_{\rm 2}$~(s)} &
\colhead {$t_{\rm 3}$~(s)} &
\\}
\startdata
&&&C\&O shell burning&&& \\
\hline
no--rot      & 3.46  &  36.94	& 0.	  & 136 &  - 	&   545    &  749  &  1000    \\  
rot--ST      & 3.46  &  36.94	& 5.5   & 368	  &  3957   &	545    &  749  &  1000    \\  
rot--2	     & 3.46  &  36.94	 & 185.6 & 142 &  126     &	545    &  749 &   1000    \\   
\hline
&&& Si shell burning&&& \\
\hline
no--rot     & 1.64  &  36.00    & 0.	&  445   &  -	  &   300    &   400 &   500   \\   
rot--ST     & 1.64  &  36.00    & 7.1  &  362  &  1505   &   300	 &   400 &   500   \\		 
rot--2	    & 1.64  &  36.00	& 110.0  & 449 &  138     &   300	 &   400 &   500   \\	
\enddata 
\tablecomments{``no-rot'': non-rotating model. ``rot-ST'': rotating model that includes
the magnetic field effects of the Spruit-Taylor dynamo. ``rot-2'': rotating model produced
by introducing a rotational velocity profile to ``no-rot'' upon mapping to {\it FLASH}.
The rotational time-scale, $\tau_{\rm rot}$, corresponds to the rotational period (one
full revolution around the rotation axis) for the given rotational speeds in the center
of the convective shell.
\label{T1}}
\end{deluxetable*}

\begin{figure*}       
\centerline{
\hskip -0.2 in
\psfig{figure=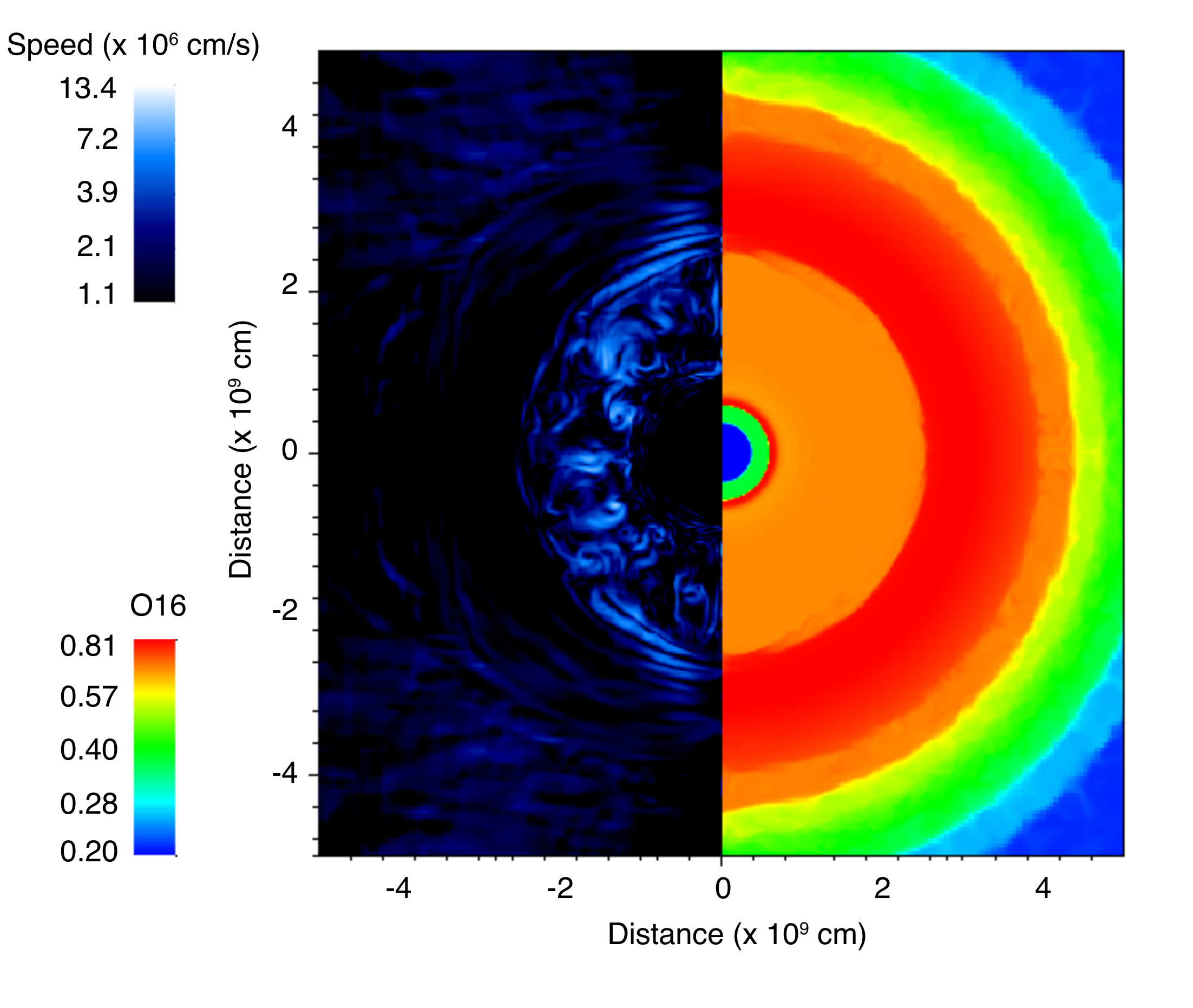,angle=0,width=2.35in}
\psfig{figure=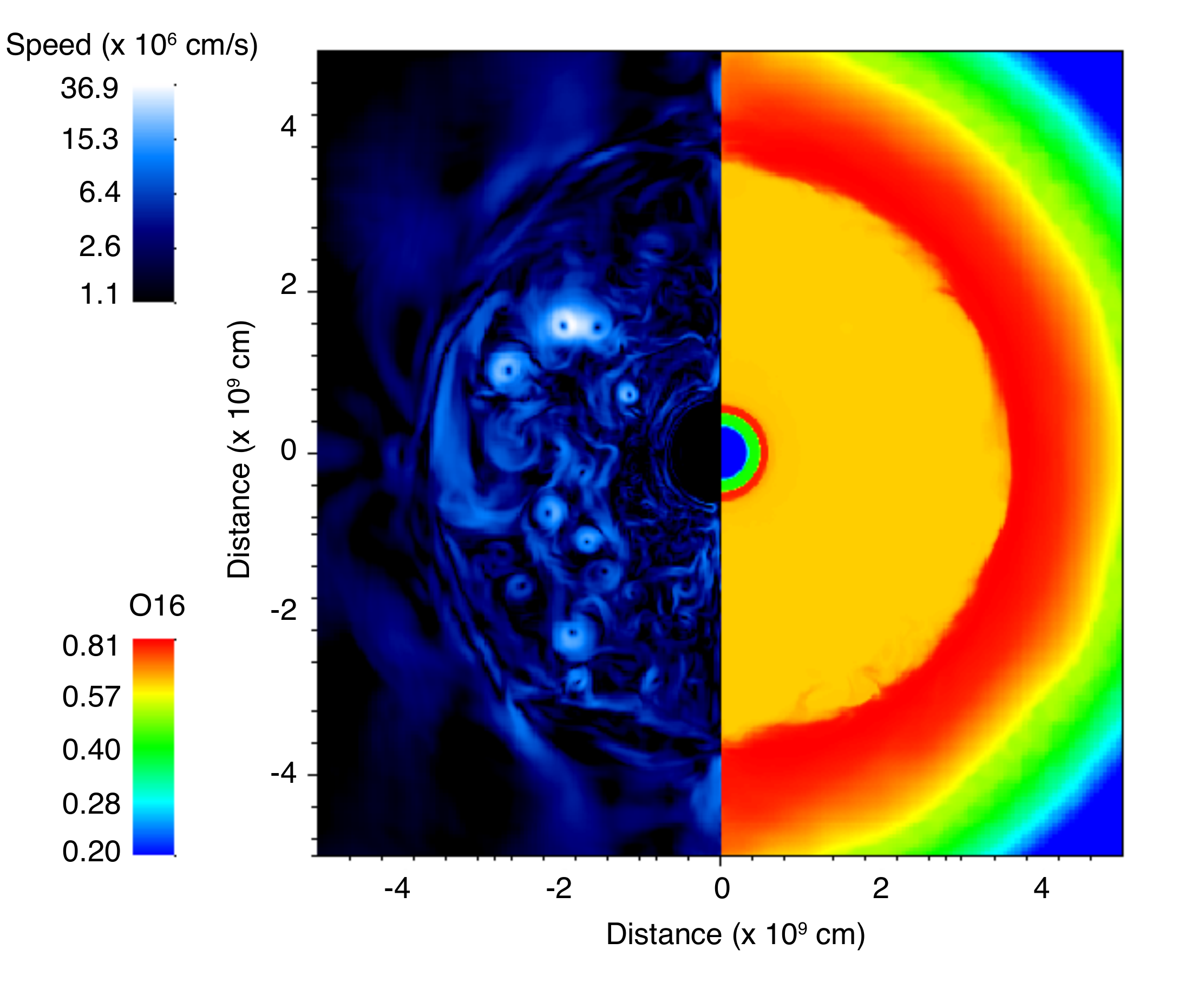,angle=0,width=2.35in}
\psfig{figure=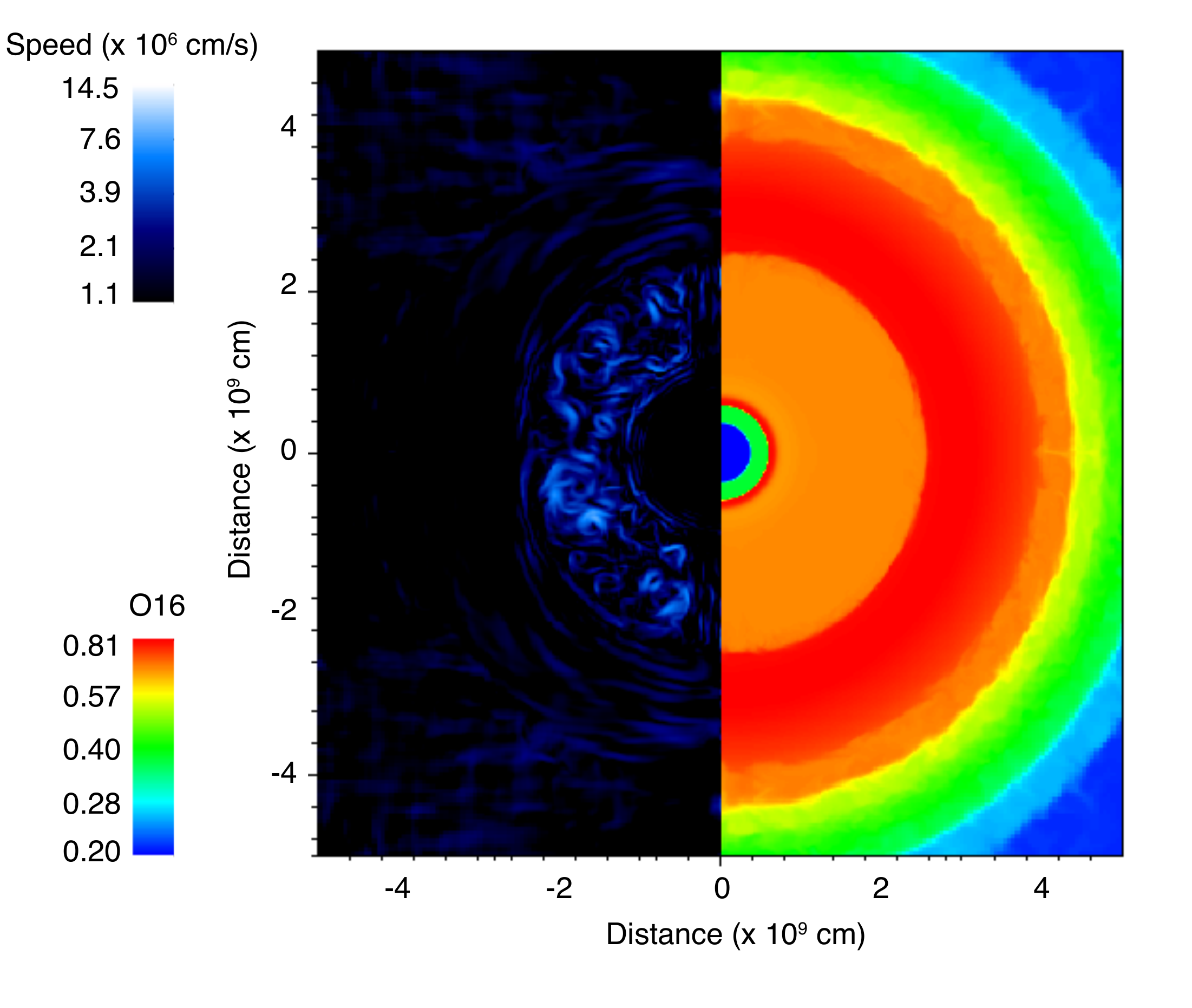,angle=0,width=2.35in}
}
\caption{Velocity magnitude and $^{16}$O mass fraction for the 
``no--rot'' ({\it left panel}), ``rot--ST'' ({\it right panel}) and ``rot--2'' ({\it right panel}) C-- \& O--shell
burning models at the end of the simulation ($t = t_{\rm 3}$).\label{Fig:coshellsims}}
\end{figure*}

\begin{figure*}       
\centerline{
\hskip -0.2 in
\psfig{figure=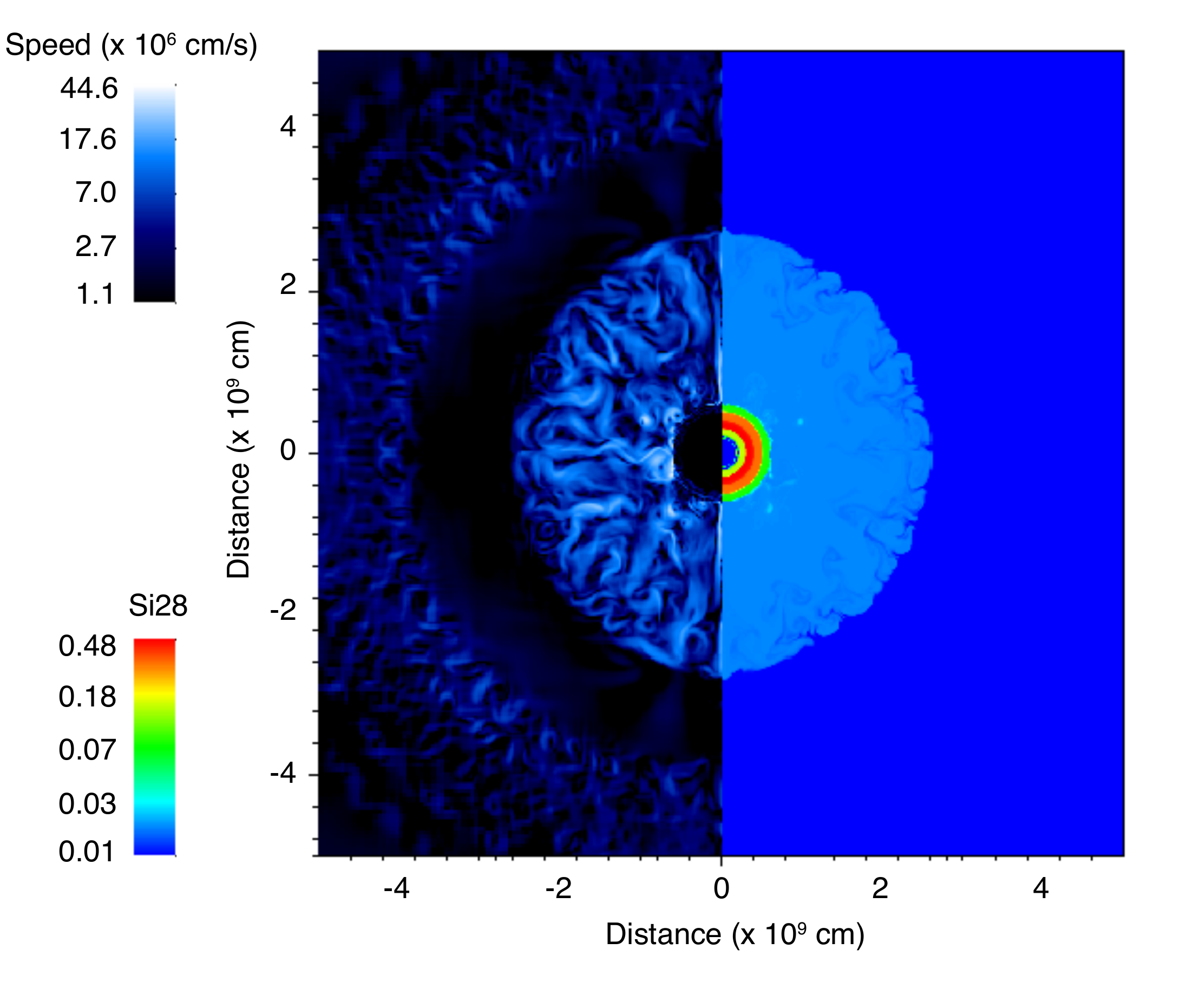,angle=0,width=2.35in}
\psfig{figure=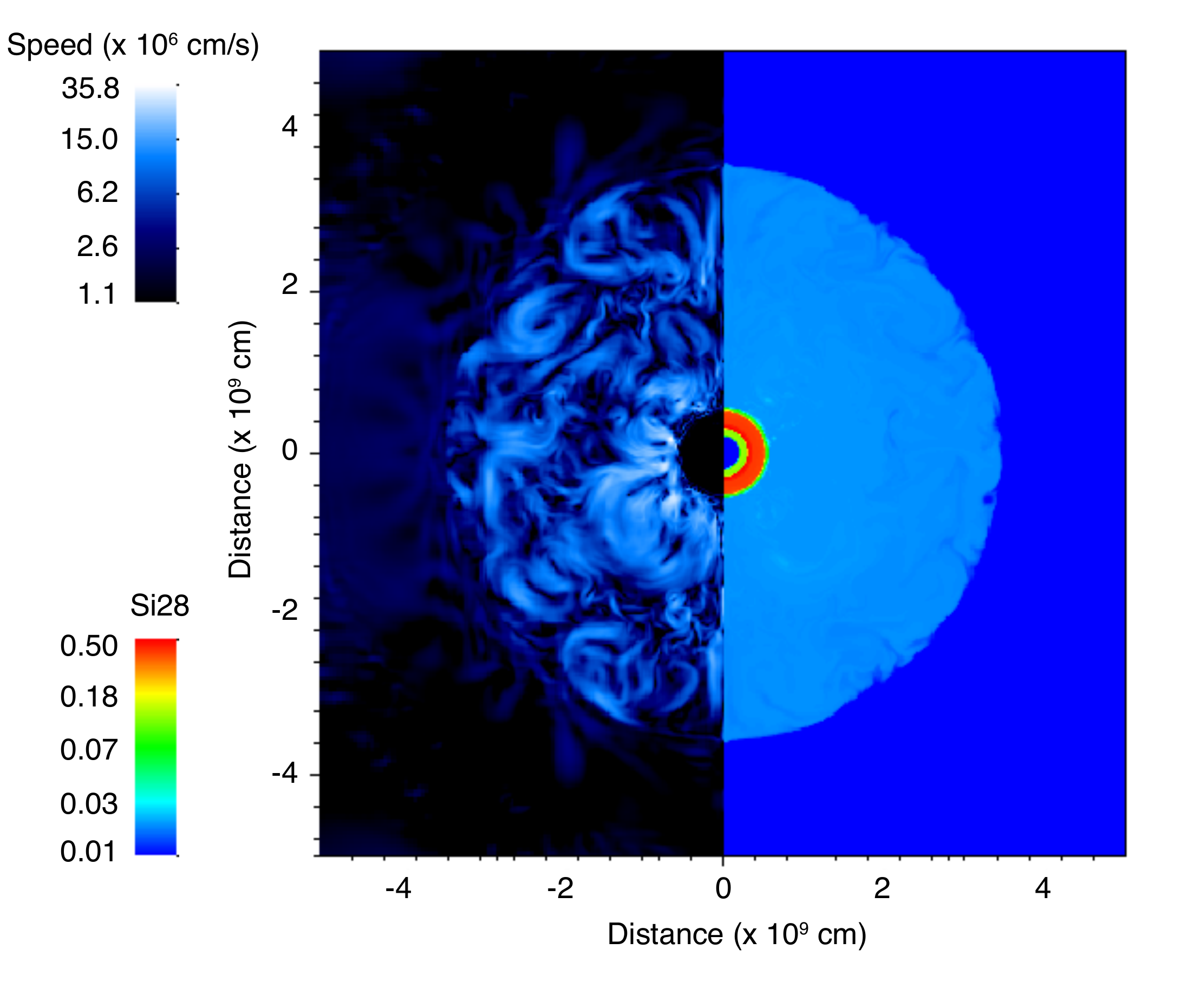,angle=0,width=2.35in}
\psfig{figure=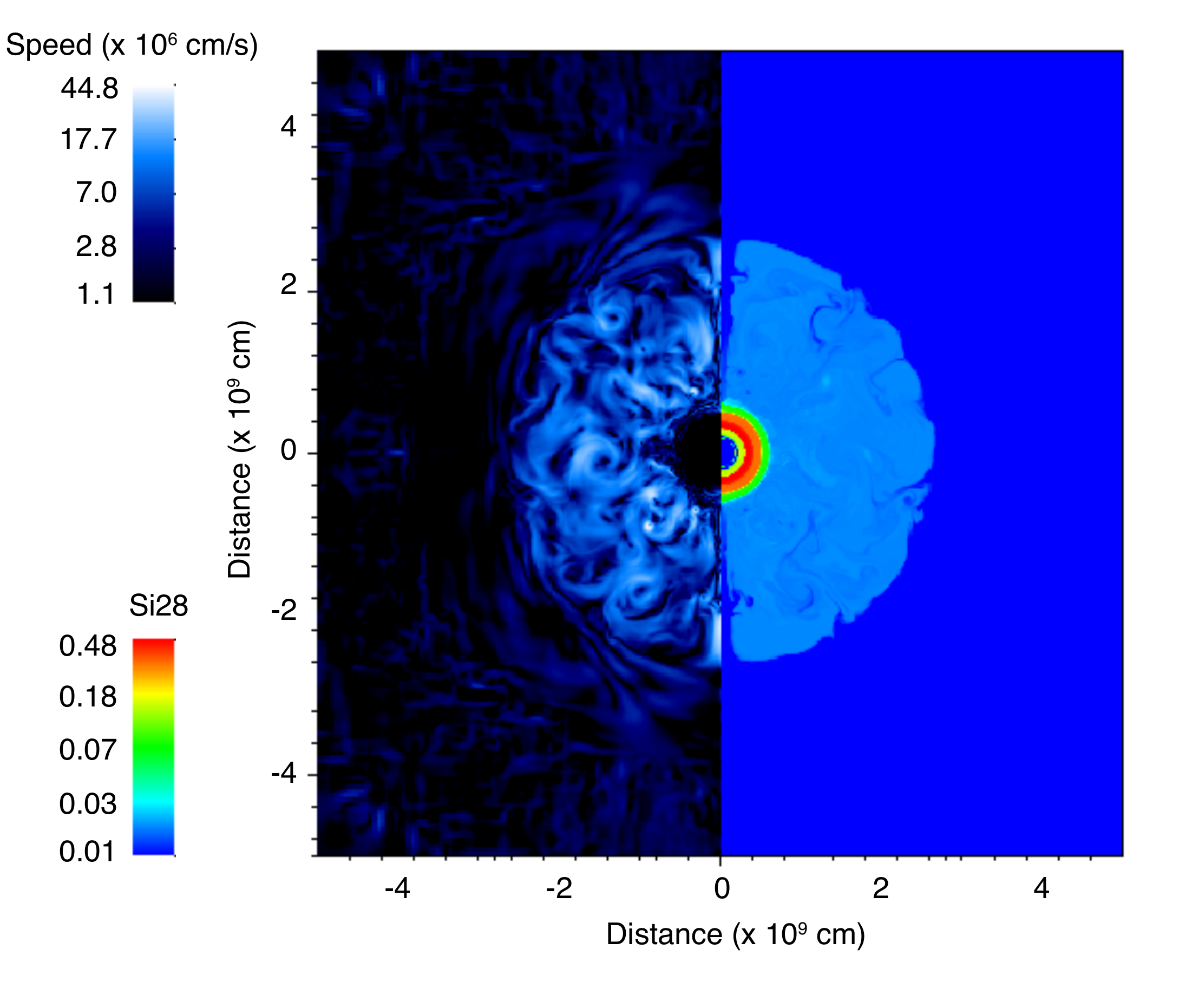,angle=0,width=2.35in}
}
\caption{Velocity magnitude and $^{28}$Si mass fraction for the 
``no--rot'' ({\it left panel}), ``rot--ST'' ({\it right panel}) and ``rot--2'' ({\it right panel}) Si--shell
burning models at the end of the simulation ($t = t_{\rm 3}$).\label{Fig:sishellsims}}
\end{figure*}

\subsection{{\it 2D FLASH hydrodynamics.}}\label{Sec:2dFLASH}

The six 1D {\it MESA} models imported into {\it FLASH} are listed in Table~\ref{T1}. The C \& O--shell burning {\it FLASH}
models are run for 1000~sec (simulated time), and the Si--shell burning models for 500~sec.
The dynamical effects of rotation and angular
momentum conservation are handled by the unsplit piecewise parabolic method (PPM) hydrodynamic solver 
in {\it FLASH} \citep{2009ASPC..406..243L}. We note that the newest implementation of the unsplit solver
in {\it FLASH} handles species advection in a way that is nearly identical to consistent multifliud advection
scheme methods \citep{1999A&A...342..179P}.
The 1D rotational velocity profiles were mapped on the 2D grid assuming ``shellular'' rotation as 
vectors with direction perpendicular to the $R$--$z$ plane
of the simulation and the rotation axis coincident with the polar axis. In this approximation the angular velocity, $\Omega(r)$,
is constant for a particular {\it spherical} shell or on equipotential surfaces \citep{1992A&A...265..115Z,1997A&A...321..465M}.
More specifically, each grid cell was given a rotational velocity 
$v_{\rm rot} = \Omega(r) R$ where $\Omega(r) = v_{\rm rot}(r)/r$
with $v_{\rm rot}(r)$ being {\it MESA} 1D rotational velocity and $r$ the spherical radial coordinate ($r = (x^{2}+y^{2})^{0.5}$). 

The transition from 1D {\it MESA} to 2D {\it FLASH} was smooth in terms of the important physics involved; a hardcoded
21--isotope' network identical to the one used in {\it MESA} was employed \citep{2015ApJ...808L..21C} 
and the ``Helmholtz'' EOS was used. The main inconsistency is the treatment of convection in transition from 1D to 2D: 
in 1D the convective energy transport and cascade is treated via approximate MLT prescriptions, while in 2D and 3D
convective flow naturally develops in the fluid within unstable regions.
The mapping from 1D to 2D triggers an initial dynamical transient that lasts for a $\sim$~100--300~sec before 
a quasi-steady state
is re--established. For our VSH analysis presented in \S~\ref{VSHdecomp} we use the {\it FLASH} output (``snapshots'')
at three different times, well after ($>$~300~sec for C \& O--shell burning and $>$~200~sec for Si--shell burning) 
the initial transient has transversed the computational domain. 
Table~\ref{T1} also details the properties of the convective shells and simulation output. 

All 2D {\it FLASH} simulations were run on the Texas Advanced Computing Center {\it Stampede} supercomputer.
The size of all simulation domains was chosen to be $10^{10}$~cm including both the core and the convective shells
of all models. The maximum resolution chosen was 9~km corresponding
to convergence in total energy and mass at a $\sim 10^{-7}$~level over the course of the simulation. 
At that resolution we are able to resolve Eddy sizes covering a considerable
range of the turbulent energy cascade (from $\sim$~30,000~km down to $\sim$~10~km). 
We should caution, however, that the turbulent energy cascade in 2D is inherently inverted 
\citep{1994ApJS...93..309P,2014ApJ...786...83T,2015ApJ...799....5C} 
and full 3D treatment is required to accuretly reproduce it. 
The morphology of the flow changes significantly from 2D to 3D, and the velocity scale is moderately higher in 2D \citep{2007ApJ...667..448M}.
We consider our simulations as an initial, exploratory step. Each simulation was
run on 128 cores and the wallclock time ranged from 14 to 22 hours for a total of $>$~12,000 core--hours
used. Figures~\ref{Fig:coshellsims} and~\ref{Fig:sishellsims} show the O$^{16}$ or Si$^{28}$ mass fraction and the
2D ($x$ and $y$) velocity magnitude at the end of each simulation. The prevalence of convection is apparent in all cases with
characteristic velocities reaching $\sim$~150~km~s$^{-1}$ for C \& O--shell burning and $\sim$~450~km~s$^{-1}$ for Si--shell burning in
the ``no--rot'' and ``rot--2'' cases. It is noteworthy that by the end of all simulations the velocities in the ``rot--ST'' cases
were much higher for C \& O--shell burning ($\sim$~370~km~s$^{-1}$) but slightly lower for Si--shell burning ($\sim$~360~km~s$^{-1}$).

The convective elements seen in the velocity magnitude panels span a range of sizes, with dominant large--scale motions that cover
more than half of the size of the convective shells, as well as smaller scale vortices which are just visible (for example in the ``rot--ST'' C \& O--burning shell).
Convection is established $\sim$~200-300~sec after the start of each simulation, after the initial transient exits the computational domain. The
evolution past 200~sec shows large scale convective currents that cover the entire shell, breaking down to smaller structures and high velocity
vortices interacting with each other. Mild mixing is also seen, with instabilities developing at the interfaces of the convective
shells. This mixing can be due to the process of turbulent entrainment also seen by \citet{2007ApJ...667..448M}. 
In all simulations, convective elements interact with the inner (core) boundary and break into smaller structures that then subsequently
reunite while rising upwards. Artifical flows are seen near the axis of the simulation ($\sim$~6~deg), 
a common issue of 2D cylindrical treatment. 

\section{VSH DECOMPOSITION OF CONVECTIVE SHELLS}\label{VSHdecomp}

Output from the {\it FLASH} simulation is taken at three instances, $t_{\rm 1}$, $t_{\rm 2}$ and $t_{\rm 3}$ (the end of the simulation); all are given in Table~\ref{T1}. 
A total of 18 snapshots for all 6 cases, are post--processed using the VSH analysis implemented in the code. 
These time--scales are chosen to represent different evolution
phases during shell convection, well after the initial dynamic transient and several 
convective turnover time--scales after that ($\simeq$~10-100 for C \& O--shell burning and $\simeq$~10-50 for Si--shell burning).

The goal of the VSH analysis is to decompose the momentum density field within the selected shell into radial 
($A_{\rm nlm}$) and solenoidal ($B_{\rm nlm}$ and $C_{\rm nlm}$)
modes and then calculate the power spectrum for each of the modes to determine the global properties of 2D or 3D fluid motion. In our
case we analyze a 2D momentum density field, therefore the solenoidal $B_{\rm nlm}$ modes are irrelevant because they cancel out
and will not be discussed further.
Also, because our simulations were conducted in a full domain possessing reflection symmetry, we expect the presence of an odd-even effect in the power
spectra, a feature inherent to the 2D treatment. 
The first step is to determine the radial limits of the convective regions where VSH is applied. For this,
volume--weighted radial momentum density profiles are calculated (Equation 22 of \citealt{2014ApJ...795...92C}). Locations
where the radial component reaches a minimum are used as our final choice for $R_{\rm in,sh}$ and $R_{\rm out,sh}$; these are also presented in Table~\ref{T1}.

The next step is to declare the maximum radial and angular
resolution for the VSH components, both of which can be expressed as a length scale, $\lambda_{\rm r}$. 
The chosen resolution scale then determines the number of radial ($n$) and angular ($l$ and $m$) modes required for the expansion in the VSH components, which
are given by the following formulae:

\begin{equation}
n_{\mathrm{max}} = \frac{2(R_{2}-R_{1})}{\lambda_{\rm r}},\label{eq:nmax}
\end{equation}
and
\begin{equation}
l_{\mathrm{max}} = \frac{\pi(R_{1}+R_{2})}{2 \lambda_{\rm r}},\label{eq:lmax}
\end{equation}
where the total number of modes in 2D momentum density field decomposition is
\begin{equation}
N_{\mathrm{total}} = (n_{\mathrm{max}}+1)(l_{\mathrm{max}}+1).\label{eq:ntotal}
\end{equation}

For the C \& O--shell burning models we choose $\lambda_{\rm r} = 1.585 \times 10^{8}$~cm, corresponding to $l_{\rm max} =$~40 and $n_{\rm max} =$~42,
while for the Si--shell burning models we choose  $\lambda_{\rm r} = 1.495 \times 10^{8}$~cm, $l_{\rm max} =$~40 and $n_{\rm max} =$~46. We choose to truncate
radial $n>$~20 modes because we find that their contribution to the total power is minimal ($10^{-5} - 10^{-4}$~level) and we can thus reduce computation time.
With these choices, a total of 861 modes were caclulated for each of the 18 {\it FLASH} snapshots. 
For the purposes of our study, 
we calculate reduced VSH spectra by firstly summing over all the ``phase'' ($m$) components. Then, we calculate the reduced angular ($l$) and radial $n$ power
spectra by summing over either $n$ or $l$, respectively, for instance
$\alpha_{l}^\prime\equiv\sum_{n} \alpha_{nl}$, $\alpha_{n}^{\prime}\equiv\sum_{l} \alpha_{nl}$, and similarly for the solenoidal modes. In our presentation
of VSH power spectra later, we will simply refer to the radial and
solenoidal modes as $A$ and $C$, respectively (for the exact definitions of $A$ and $C$ consult Equations 6-8 of \citealt{2014ApJ...795...92C}).

\begin{figure*}       
\hskip -0.03 in
\centerline{\psfig{figure=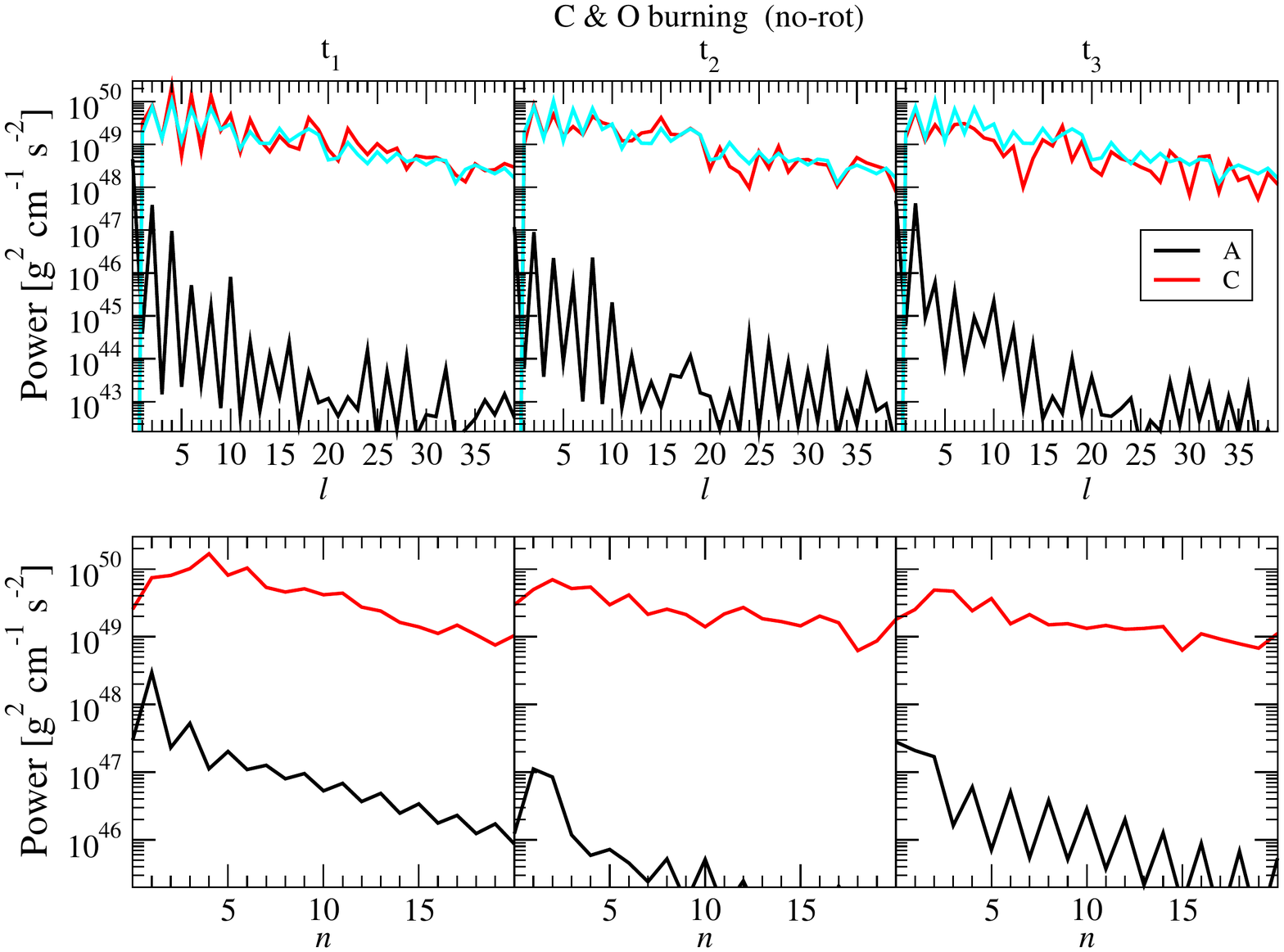,angle=0,width=12cm,height=8cm,trim=0.3in 0.8in 0.5in 0.25in,clip}} \\
\hskip 1.0 in
\centerline{\psfig{figure=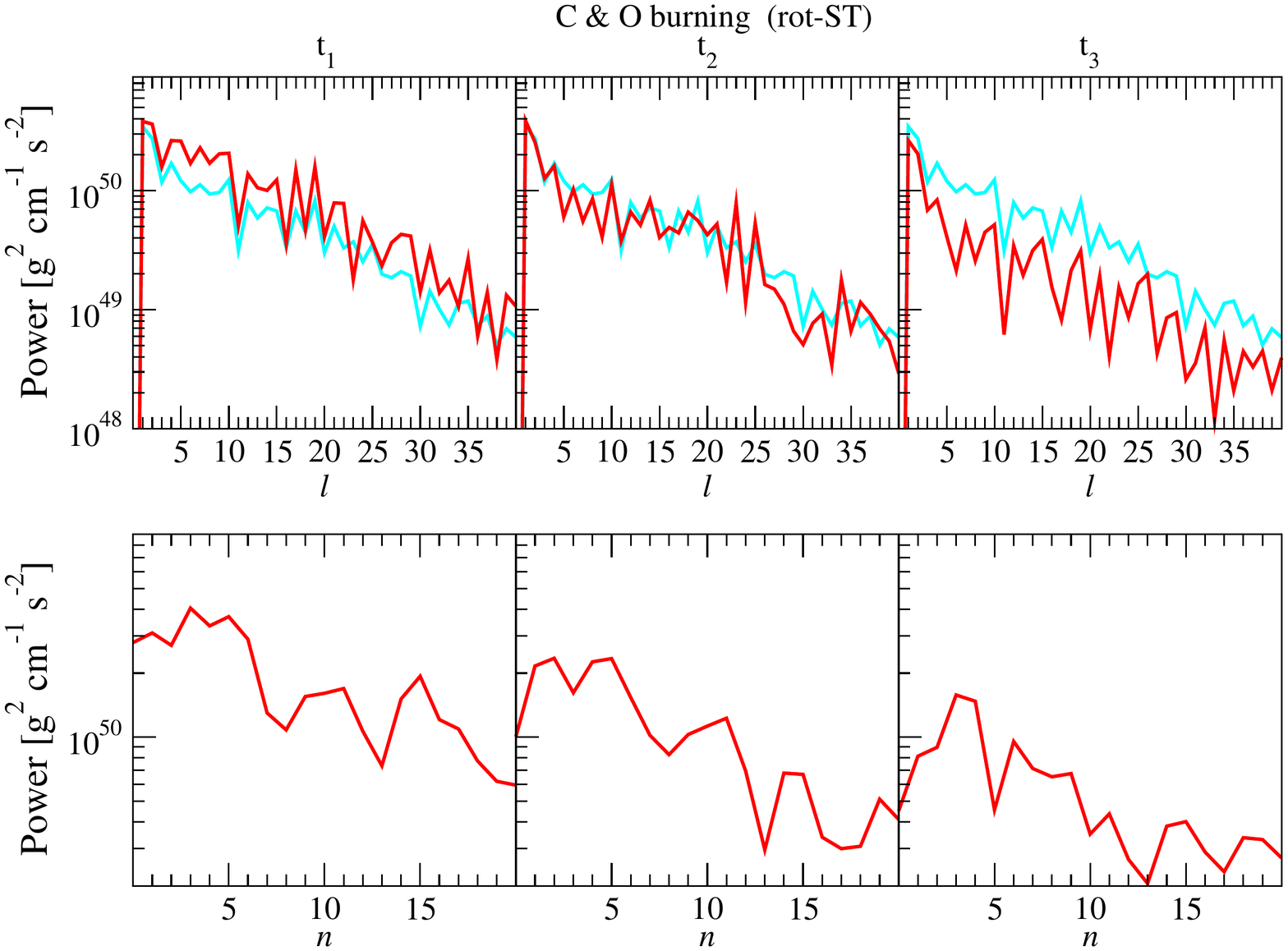,angle=0,width=12cm,height=8cm,trim=0.3in 0.8in 0.5in 0.25in,clip}} \\
\hskip 1.0 in
\centerline{\psfig{figure=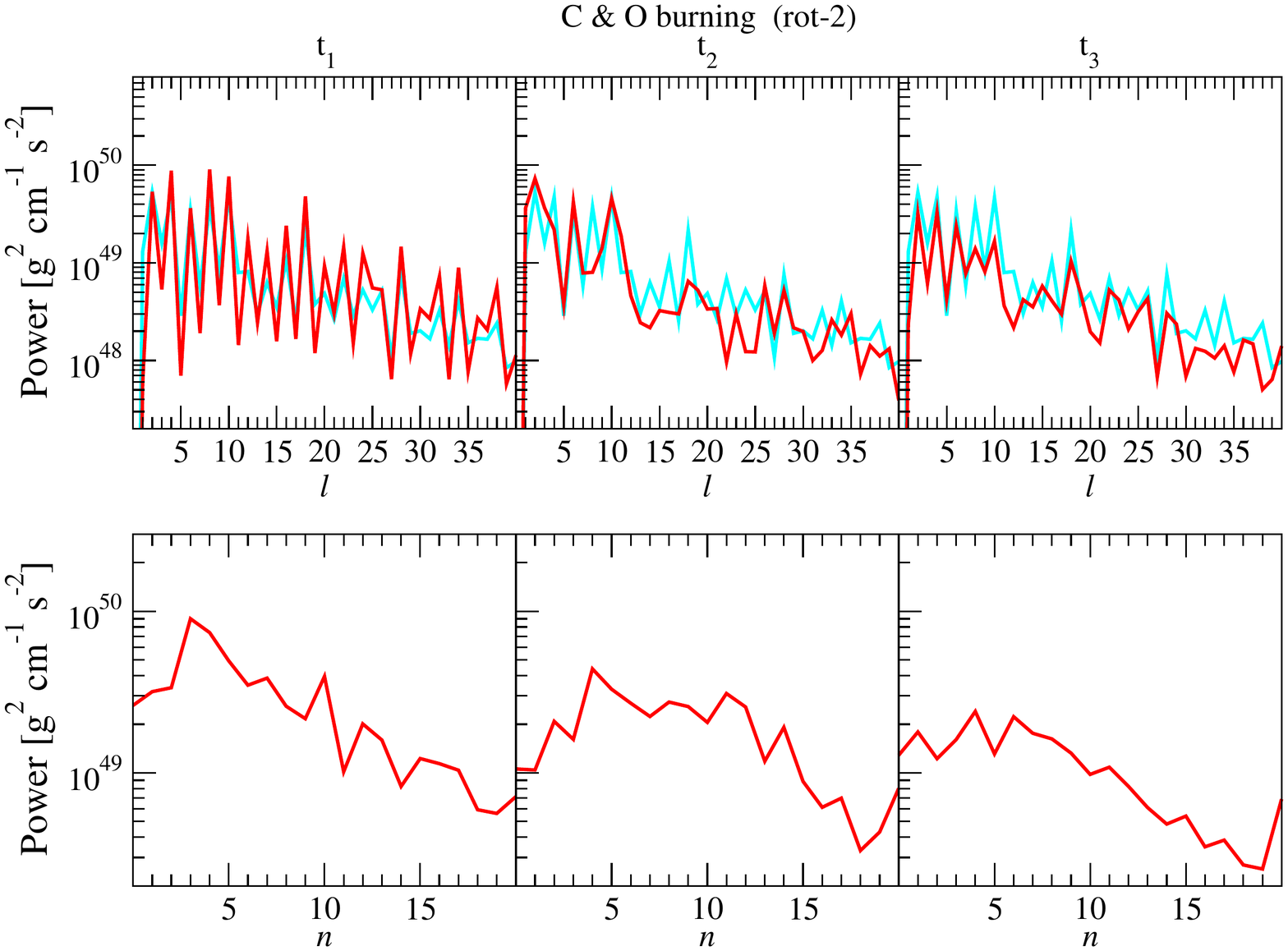,angle=0,width=12cm,height=8cm,trim=0.3in 0.8in 0.5in 0.25in,clip}}
\caption{Evolution of reduced VSH power spectra in {\it l} 
and {\it n} for the ``no--rot'' ({\it upper panel}), ``rot--ST'' ({\it middle panel}) and ``rot--2'' ({\it lower panel}) 
C \& O--shell burning models.  
$A$ (irrotational) and $C$ (solenoidal) modes are shown with black and red curves respectively. Since the $A << C$
always, we show the power in the irrotational modes only for the ``no--rot'' case. The cyan curves show time--averaged
$C$ spectra for the three snapshots.\label{Fig:o_vshps}}
\end{figure*}

\begin{figure*}       
\hskip -0.03 in
\centerline{\psfig{figure=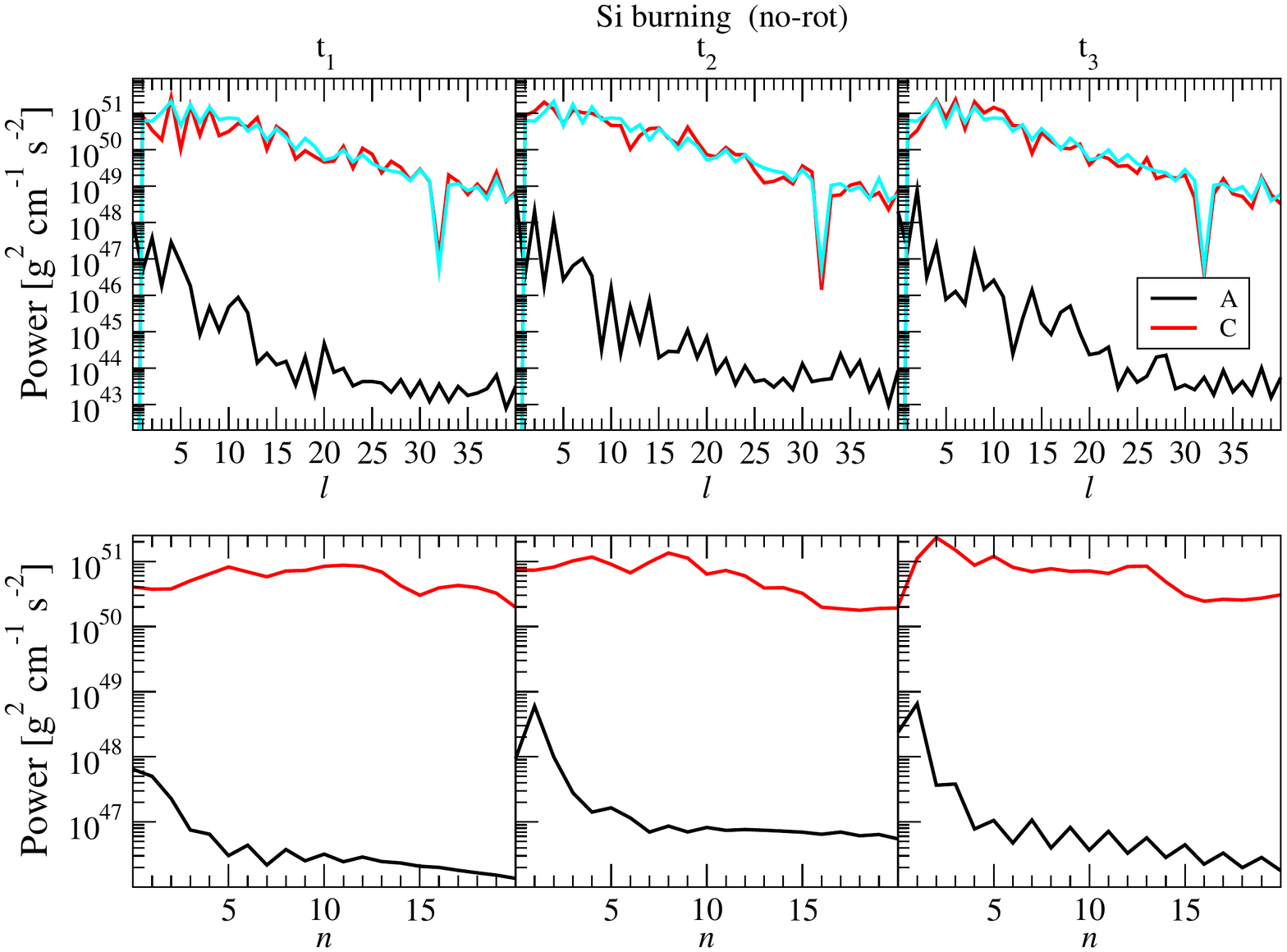,angle=0,width=12cm,height=8cm,trim=0.3in 0.8in 0.5in 0.25in,clip}} \\
\hskip 1.0 in
\centerline{\psfig{figure=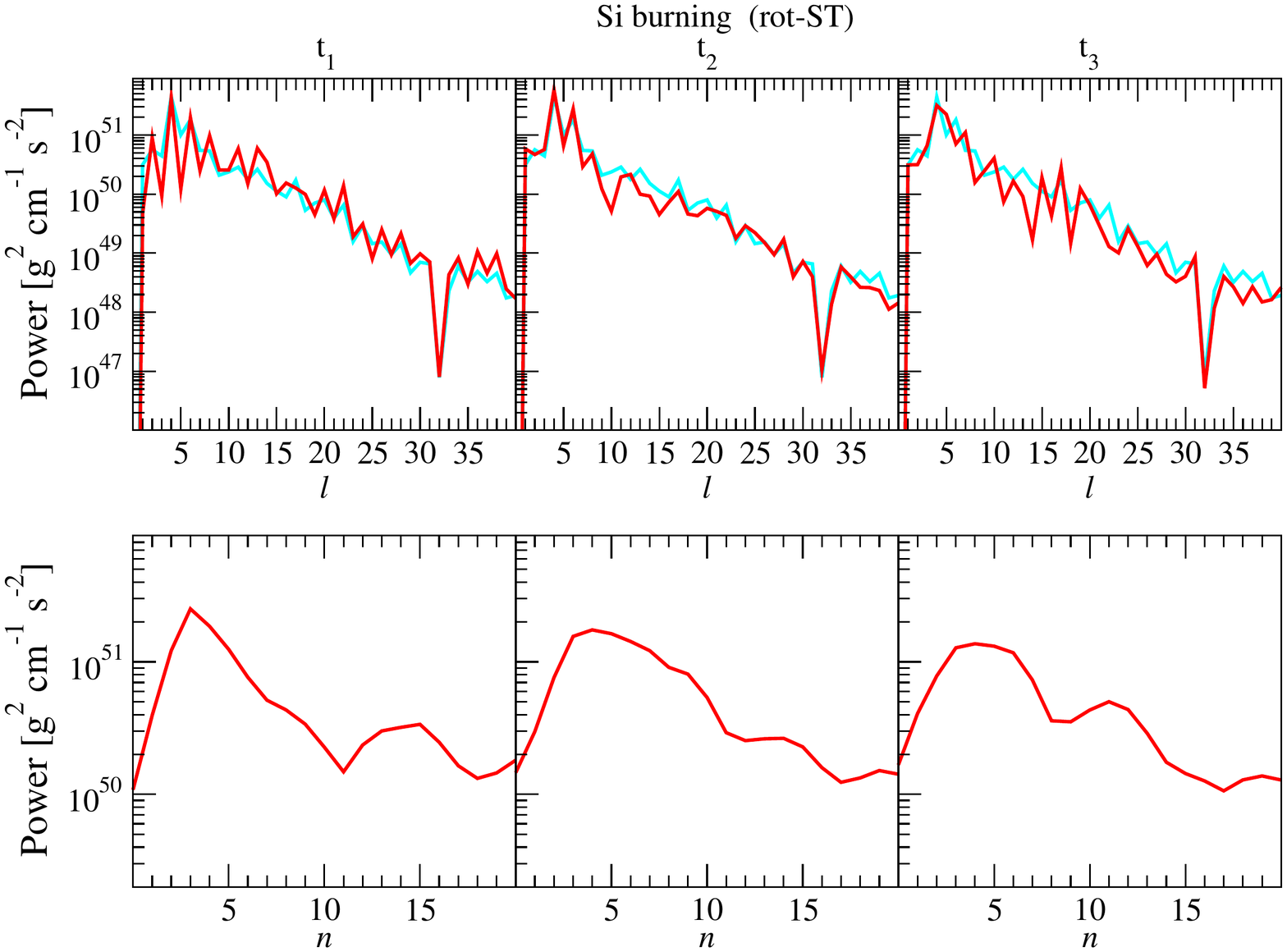,angle=0,width=12cm,height=8cm,trim=0.3in 0.8in 0.5in 0.25in,clip}} \\
\hskip 1.0 in
\centerline{\psfig{figure=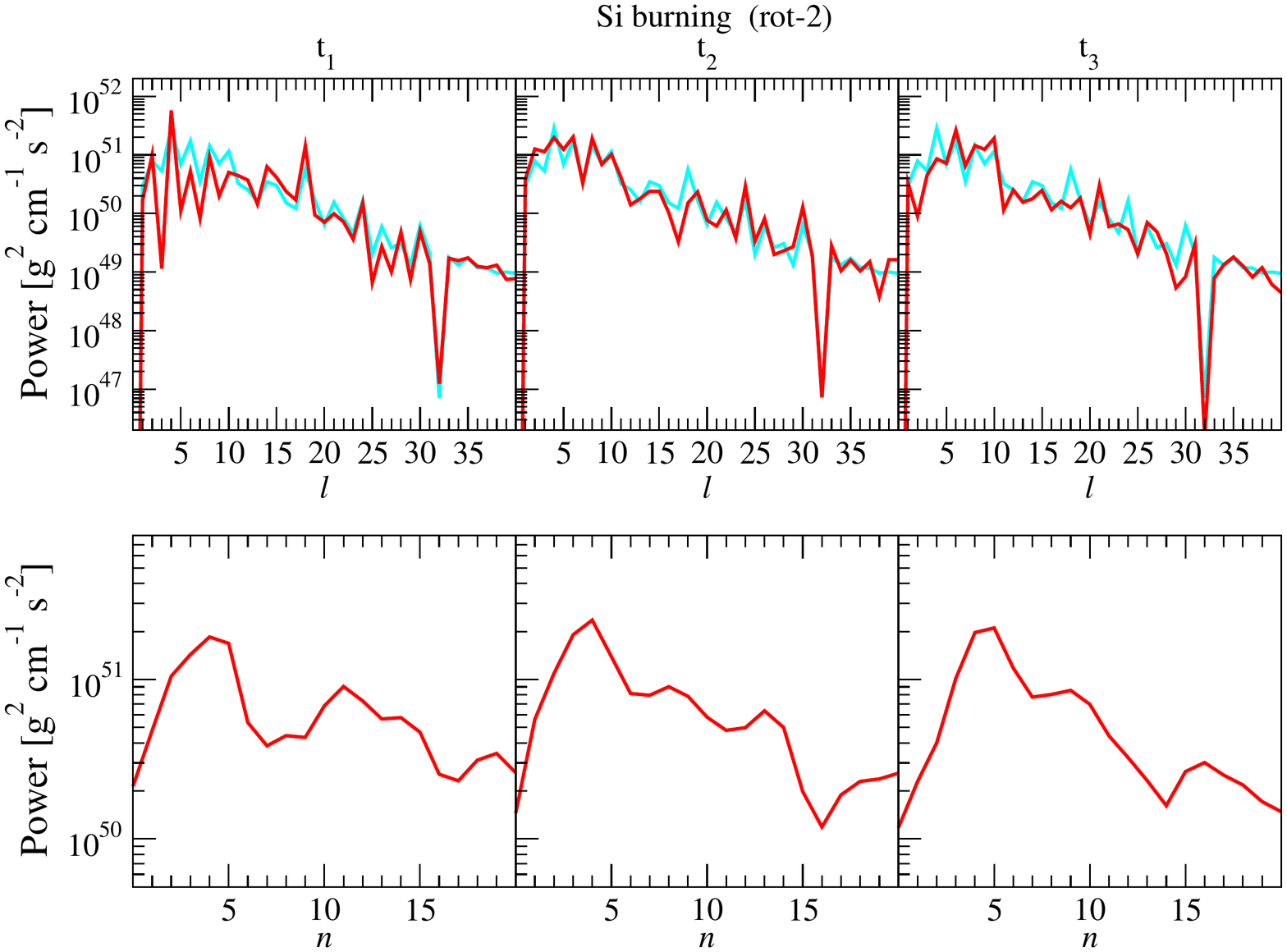,angle=0,width=12cm,height=8cm,trim=0.3in 0.8in 0.5in 0.25in,clip}}
\caption{Same as Figure~\ref{Fig:o_vshps} but for the Si--shell burning models.\label{Fig:si_vshps}}
\end{figure*}

\begin{figure*}
\begin{center}
\includegraphics[angle=0,width=15cm,trim=0.0in 0.25in 0.5in 0.15in,clip]{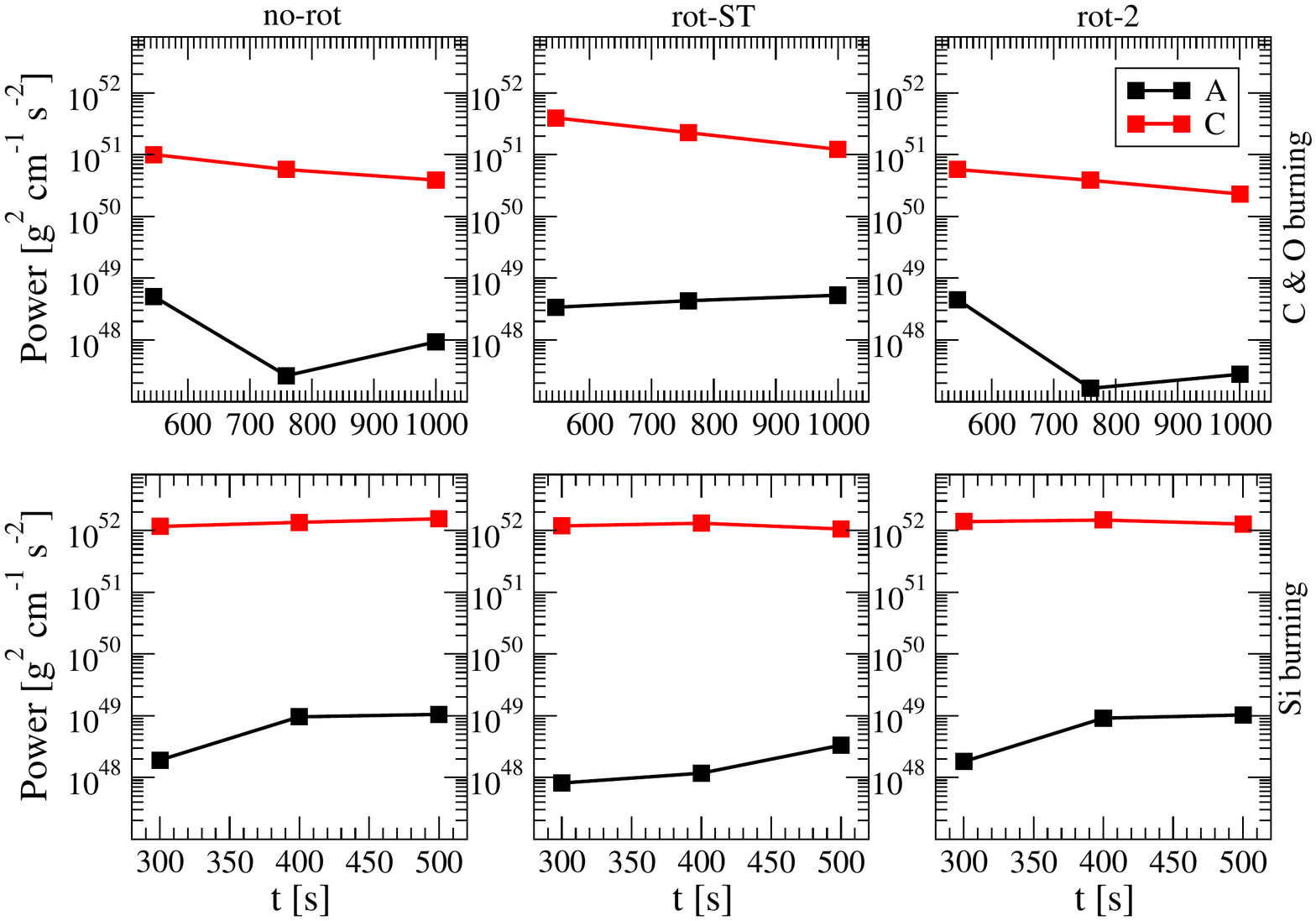}
\caption{Evolution of total VSH power in irrotational modes ($A$; black curves) and solenoidal
modes ($C$; red curves) for all the progenitor models of Table~\ref{T1}. The C\&O shell
burning models are shown in the upper panels and the Si shell burning models in the lower
panels.
\label{Fig:total_power_evol}}
\end{center}
\end{figure*}

\subsection{{\it Shell O-- and C--burning.}}\label{COburnVSH}

Figure~\ref{Fig:o_vshps} shows the reduced VSH power spectra for all models undergoing C\&O--shell burning. 
The evolution of the total power, summed over all components, is shown in the top row of Figure~\ref{Fig:total_power_evol}. For all cases the bulk of
convective power is concentrated in large scales ($l <$~10, $n <$~5), with the peak values implying angular scales of $\sim$~$1.6 - 3.2 \times 10^{9}$~cm
and radial scales of approximately the same range, revealing a nearly circular characteristic shape for the convective eddies. These eddies can be comparable in size to the shell itself. 
For some spectra, a secondary peak of power is observed at smaller values of $l$ ($\sim$~15-20), indicating that a small fraction of the total power 
is possessed by smaller scales ($\sim 4.3 \times 10^{8}$~cm). 
The overall slope of the spectra remains nearly fixed over time. 
As expected for a convective velocity field confined in a shell,
the power in the solenoidal modes is clearly dominant over that in the radial modes. 

Regardless of the degree of rotation, the total power in dominant solenoidal modes 
declines over the course of the simulations as the convective energy cascade settles. A comparison between the 
``no--rot'' and ``rot--2'' case shows that the addition of rotation in otherwise equal stellar structure has little effect on the global properties of
convection. A small reduction of total power is apparent, possibly related to the effects of centrifugal forces and the existence of an extra degree of freedom
(movement perpendicular to the simulation domain). On the other hand, the algorithm for
calculating the effects of rotation during stellar evolution leads
to a pre--SN star with clearly stronger convection during the C\&O--shell burning phase (``rot--ST'' model, upper middle panel in Figure~\ref{Fig:total_power_evol}).
Indeed, the final peak 2D-velocity magnitude in the ``rot--ST'' model is more than double those of the ``no--rot'' and ``rot--2'' models. We discuss
this in more detail in \S~\ref{Disc}.

\subsection{{\it Shell Si--burning.}}\label{SiburnVSH}

The VSH power spectra for the cases of shell Si--burning are shown in Figure~\ref{Fig:si_vshps}. The evolution
of the total power is shown in the bottom row of Figure~\ref{Fig:total_power_evol}. The reduced spectra reveal that large scales also dominate over smaller
scales during shell Si--burning, with $l$ peaking in the range 4--6 throughout the evolution. The corresponding angular length scales range from $1.5 \times 10^{9}$~cm
to $2.0 \times 10^{9}$~cm. The radial scales are also in the same range implying nearly circular shape for the convective elements, which 
are at about half
the size of the convective shell. Secondary peaks occur at smaller scales throughout the VSH spectral evolution ($\sim 10^{9}$~cm). As in the case of C\&O--shell
burning, the spectral slopes remain consistent over time and solenoidal models dominate radial models by a factor of $\sim$~10,000 for $l <$~5.

The evolution of the total power in the solenoidal ($C$) modes 
shows that convection during Si--shell burning is about 10 times stronger than convection
during C\&O--shell burning, as expected from the higher rates of local energy generation. 
The total power does not seem to vary significantly over time for all rotation rates. A small 
reduction of total
power in the $C$--modes 
is seen by the end of the simulations 
for the ``rot--ST'' and ``rot--2'' cases. For models initially identical, modulo the inclusion
of rotation (``no--rot'' versus ``rot--2''), this effect may be due to the dynamical impact of centrifugal forces and the extra degree of freedom, as argued for the case of C\&O--shell
convection. We do, however observe a qualitative difference between the two stages of shell burning: 
during Si--shell burning, the ``rot--ST'' model exhibits nearly identical and, at late times, somewhat lower
convective power as compared to the ``no--rot'' model, while the opposite behavior was observed for C\&O--shell burning. We return to this issue in the
following section.

\section{DISCUSSION AND CONCLUSIONS}\label{Disc}

In this paper we explored the effects rotation on the convective properties of a 20~$M_{\odot}$ pre--SN progenitor star, more specifically during C\&O--shell burning and Si--shell burning in the hours to months prior to core--collapse. 
We studied rotation both by including it self--consistently during the evolution of the star but also by imposing it to an othewise identical, non--rotating model in order to better 
isolate its effects. 
We run 2D {\it FLASH} simulations with the rotational velocity field mapped and pointing inwards perpendicular to plane of the simulation domain.  
We employed vector spherical harmonic decomposition of the momentum density field to characterize convection at different times during the hydrodynamic simulations.\ Our initial exploration of the influence of rotation on the properties of convection deep in the core of a massive star can (and should) be improved by using 3D simulations, where the turbulent cascade goes from large scales to small scales.

We find that, regardless of the degree of rotation, the characteristic size of the convective elements is somewhat larger during the C\&O--shell burning phase compared to the Si--shell burning phase, while the characteristic convective velocities are more than twice as large during Si--shell burning as compared to C\&O--shell burning.  
The dominant scales implied by the resulting VSH power spectra span more than 50\% of the size of the convective shells, with secondary typical scales as small as $0.5 - 1 \times 10^{9}$~cm. 
The slope of the VSH power spectra stays consistent over time regardless of the presence or absence of rotation. 
In all cases we find nearly circular shapes for the characteristic convective elements. 
In terms of the 2D convective flow properties, our results are in agreement with those of \citet{2007ApJ...667..448M} (see also \citet{2015IAUS..307...20M}; \citet{2015IAUS..307...98C}).  
In particular, we find a highly intermittent flow and mixing due to turbulent entrainment.

In terms of the effects of rotation on the strength and properties of convection alone we find our 2D simulations suggest minimal impact, and generally lead to an overall small reduction of the total convective power stored in solenoidal motions (``no--rot'' versus ``rot--2'' models) regardless of the nature of convective burning (C, O, or Si shell). 
We suggest this is due to the effects of the centrifugal forces mildly expanding the star, and thus changing the locations of the convective shells over time, subjecting the shell material to lower temperatures that, in turn, trigger lower nuclear burning rates and weaker convection. 
That alone may lead to initial velocity perturbations of smaller amplitude than in the case of zero rotation which may have a small impact on the susceptibility to a successful SN explosion following iron core collapse \citep{2013ApJ...778L...7C}.

We emphasize that, to answer the question of whether the inclusion of rotation significantly changes the ICs to the core--collapse SN mechanism, self--consistent evolution with the effects of rotation included is the proper approach to take. 
In our analysis we have followed the evolution of a 20~$M_{\odot}$ SN progenitor with both the effects of rotation and magnetic fields included in the transport of angular momentum and chemical mixing using the {\it MESA} code (``rot--ST'' models).
We find differences in the sign of the effect depending on the nature of convective shell--burning: during C\&O--shell burning there is more power
stored in the solenoidal components than in the case of no rotation while during Si--shell burning the effects are very small with hints of even reduction of
the solenoidal mode power by the end of the simulation.
We attribute this effect to differences in the initial {\it MESA} models for the two distinct shell--burning
stages. 
A careful look in the upper panel of Figure~\ref{Fig:CO_characteristics} shows that the ``rot--ST'' $\epsilon_{\rm nuc}$ profile
during C\&O--shell burning has a secondary peak (due to O--burning) that is nearly an order of magnitude greater than the corresponding one for the ``no--rot'' model. 
On the contrary, in Figure~\ref{Fig:Si_characteristics} we see that during Si--shell burning the peaks in the $\epsilon_{\rm nuc}$ profile for the ``no--rot'' model in the region $4 - 6 \times 10^{8}$~cm are greater than the corresponding ones for ``rot--ST''. 
This is due to the enhanced chemical mixing by
the ST mechanism during the C\&O--shell burning phase that effectively recycles fresh fuel from outer layers to deeper and hotter regions enabling faster specific nuclear energy generation rates. 
In contrast, during the later and more short--lived Si--shell burning phase the ST mechanism does not have
the same radial extent and efficiency to instigate similar effects. 
This result illustrates that the presence of efficient mixing mechanisms need to be studied self--consistently and in more detail since they can alter the convective properties and structure of massive stars prior to CCSNe quantitatively.

If the Spruit-Taylor mechanism is even roughly correct, our
simulations suggest that the cores of most massive stars do not rotate
rapidly enough for rotation to be dynamically relevant to the CCSN mechanism (see also \citealt{2005ApJ...626..350H,2015arXiv151101471G}).  
The inclusion of rotation and attendant angular momentum-transporting
instabilities in the  stellar evolution calculation, however, does
significantly impact the nature of the convection surrounding the
pre-collapse iron core.  
This could have important implications for the CCSN mechanism itself following core collapse.
Recently, \citet{2015arXiv151200838M} suggested that above a certain rotational thershold, the magneto-rotational instability (MRI) can drive an inverse cascade of the magnetic energy generating a large--scale magnetic field that can provide the conditions for Gamma-Ray Burst jets and explain the origins of Type Ib/c SNe as well as some superluminous supernovae (SLSN) powered by the spin--down of newly--born magnetars. 
In the limit of slow and typical rotation rates explored here we do not expect MRI--induced turbulence to have an important effect on the progenitor properties for single--star evolution.  
Binary evolution seems to offer an alternative channel, with the possibility of rapidly rotating core collapse in which the MRI may be effective as an explosive and jet--forming mechanism.


\acknowledgments

We thank J. Craig Wheeler for useful conversations.
EC thanks the Enrico Fermi Institute for its support via the Enrico
Fermi Fellowship.
The authors acknowledge the Texas Advanced Computing Center (TACC) at The University of Texas at Austin for providing HPC, visualization, and storage resources that have contributed to the research results reported within this paper.
An award of computer time was provided by the Innovative and Novel Computational Impact on Theory and Experiment (INCITE) program. This research used resources of the Argonne Leadership Computing Facility, which is a DOE Office of Science User Facility supported under Contract DE-AC02-06CH11357.


\bibliography{massiverot}

\end{document}